\documentclass[aps,prb,reprint,superscriptaddress]{revtex4-2}
\usepackage{soul}
\usepackage[pdftex]{graphicx}
\usepackage{amsmath}
\usepackage{color}
\usepackage{natbib}
\usepackage{xcolor}
\begin{document}

\title{Real-space Atomic Dynamics in Liquid Gallium Studied by Inelastic Neutron Scattering}

\author{Chengyun Hua}
\email{huac@ornl.gov}
\affiliation{Materials Science and Technology Division, Oak Ridge National Laboratory, Oak Ridge, Tennessee, 37831 USA}

\author{Yadu K. Sarathchandran}
\affiliation{Department of Physics and Astronomy, The University of Tennessee, Knoxville, Tennessee, 37996, USA}

\author{Eva Zarkadoula}
\affiliation{Center for Nanophase Materials Science, Oak Ridge National Laboratory, Oak Ridge, Tennessee, 37831 USA}

\author{Wojciech Dmowski}
\affiliation{Department of Materials Science and Engineering, The University of Tennessee, Knoxville, Tennessee, 37996 USA}

\author{Douglas L. Abernathy}
\affiliation{Neutron Scattering Division, Oak Ridge National Laboratory, Oak Ridge, Tennessee, 37831 USA}

\author{Takeshi Egami}
\affiliation{Department of Physics and Astronomy, The University of Tennessee, Knoxville, Tennessee, 37996, USA}
\affiliation{Materials Science and Technology Division, Oak Ridge National Laboratory, Oak Ridge, Tennessee, 37831 USA}
\affiliation{Department of Materials Science and Engineering, The University of Tennessee, Knoxville, Tennessee, 37996 USA}

\author{Yuya Shinohara}
\email{shinoharay@ornl.gov}
\affiliation{Materials Science and Technology Division, Oak Ridge National Laboratory, Oak Ridge, Tennessee, 37831 USA}

\date{\today}

\begin{abstract}

Gallium is a prototypical liquid metal and has gained renewed attention due to its unique properties. Characterizing and elucidating its atomic dynamics remains elusive despite numerous studies, primarily due to the challenges of quantifying atomic-scale dynamics in liquids. Recent developments in inelastic neutron scattering enable us to measure the Van Hove correlation function that describes the real-space motion of liquid atoms. In this work, we use this approach to reveal the dynamics in gallium liquids and find the co-existence of two dynamical medium-range orders (MROs), which have a dynamical behavior distinct from that of the short-range order (SRO). We propose that these MROs are driven by global forces in the form of two density waves, as a direct consequence of the underlying competition between ionic core repulsion and valence electron cohesion. We suggest that the density wave approach is not only applicable to other metallic liquids exhibiting similar structural anomalies, but also offers a promising direction for elucidating the dynamics of complex liquids and glasses by linking electronic-state fluctuations to atomic dynamics.

\end{abstract}
\maketitle

\section{Introduction}

Room-temperature liquid metals and their alloys have attracted renewed interest due to their unique properties \cite{dickey_Gallium_PhysToday_2021}. When used as reaction solvents, liquid metals are fundamentally different from molecular and ionic liquids, providing a distinct environment for chemical reactions and additive manufacturing. Their unique combination of conductivity and fluidity makes liquid metals excellent materials for soft and stretchable electronics. Although liquid metals have long been used, as exemplified by mercury's historical applications, the origins of their physical properties remain largely elusive. For example, Gallium is widely known as metallic liquid with low melting point. In addition to its low melting point, gallium shows a higher density in the liquid state than in the solid state---a behavior also observed in water and silica \cite{gong_gallium_1991}.  Gallium exhibits an anomalous change in diffusivity as a function of temperature, a phenomenon observed in water as well. The presence of residual covalent dimers has been observed near the melting point, indicating the mixed nature of its atomic environment \cite{li_local_2017}. In fact, gallium exhibits a mixed nature of bonding---partly covalent and partly metallic \cite{gong_coexistence_1993}.

Despite its scientific importance and ongoing interest, the correlated atomic dynamics of liquid, including gallium, remains poorly understood. For example, the origins of the glass transition and the behavior of supercooled liquids continue to be active areas of research. It is sometimes assumed that high-temperature liquids behave in a gas-like manner, based on the belief that the kinetic energy overwhelms the potential energy. However, this assumption is justified only at very high temperatures.  In most ranges of temperature, atoms in liquid remain strongly correlated.  Their motions are thermally activated, with a temperature-dependent activation energy \cite{kivelson_thermodynamic_1995,iwashita_elementary_2013}. In this study, we investigate the atomic-scale dynamics in liquid gallium at room temperature and above, with a particular focus on how observed dynamics relate to the underlying bonding environment among gallium atoms.

Earlier studies suggested that the covalent and metallic character of solid $\alpha$-Ga coexists even in the liquid phase \cite{gong_coexistence_1993,nield_changes_1998,lambie_resolving_2024,yang_firstprinciples_2011}.  The existence of covalent dimers was considered to be a remnant of $\alpha$-Ga melts, and such bonds are typically regarded as very short-lived---approximately 50 fs at 1000 K according to one molecular dynamics simulation study \cite{lambie_resolving_2024}. Alternatively, the presence of locally ordered structures, \emph{i.e.}, molecular clusters---as opposed to more strictly defined dimers or covalent bonds---has also been proposed to explain the asymmetric feature of the first peak in the snapshot structure factor, $S(Q)$ \cite{mokshin_short-range_2015}. Oberle and Beck \cite{oberle_influence_1979}, as well as Hafner \emph{et al.} \cite{tsai_revisiting_2010,tsai_entropy_2011,hafner_structural_1990, hafner_structural_1992, jank_structural_1990-1,jank_structural_1990}, proposed that the interatomic pair potential in a polyvalent liquid metal, such as liquid gallium, consists of an ionic component represented by a repulsive core and long-range Friedel oscillations induced by conduction electrons. Despite their success in reproducing the asymmetric feature in $S(Q)$ of liquid gallium, it remains unclear how these two components of the interatomic potential affect the collective real-space dynamics of gallium atoms in the liquid state.

To elucidate the atomic-scale dynamics in liquid gallium, we use the Van Hove correlation function (VHF), which is a pair correlation function in real space and time. The VHF is defined as 
\begin{equation}
    G(r,t) = \frac{1}{4\pi \rho N r^2} \sum_{i,j} \delta (r - |\mathbf{r}_i(0) - \mathbf{r}_j(t)|),
\end{equation}
where $\rho$ is the average number density of atoms, $N$ the number of gallium atom in the system, $\mathbf{r}_i(t)$ the position of the $i$th atom at time $t$ and $\delta(r)$ Dirac's delta function \cite{vanHove_1954}. The VHF, $G(r,t)$, is derived through a double Fourier transform of the dynamic structure factor $S(Q,E)$, which can be experimentally determined via inelastic neutron or x-ray scattering measurements. Recent advances in time-of-flight neutron spectroscopy and high-resolution inelastic X-ray scattering have enabled the acquisition of inelastic scattering spectra over broad ranges of momentum and energy transfer---key requirements for reconstructing the VHF through Fourier transformation \cite{egami_correlated_2020}. Although inelastic and quasi-elastic scattering studies of gallium have been reported \cite{hosokawa_transverse_2009,khusnutdinoff_collective_2020}, these measurements were limited to the low-$Q$ range, where diffusion due to multiple atomic collisions and collective excitations dominates the scattering spectra. 

In this study, we focus on the atomic-scale correlations of liquid gallium atoms by analyzing the real-space correlation function derived from inelastic neutron scattering spectra in the 1–10\,\AA$^{-1}$ range. We find that two medium-range orders (MROs) are necessary to explain the real-space correlations beyond the first peak in $G(r,t)$, and that the asymmetric feature of the first peak in $S(Q)$ arises from the coexistence of these two MROs. We also report that the decay of the two MROs with time is significantly slower than that of the first peak in $G(r,t)$ at low temperatures, suggesting that the dynamics of the MROs are different from that of the short-range order (SRO). Finally, we discuss in detail a possible origin of these two MROs. A part of this work was presented in Ref.~\cite{hua_complexity_2025}.

\section{Results}
\subsection{Conversion of dynamic structure factor into Van Hove correlation function}

\begin{figure*}[t]
    \includegraphics[scale = 0.35]{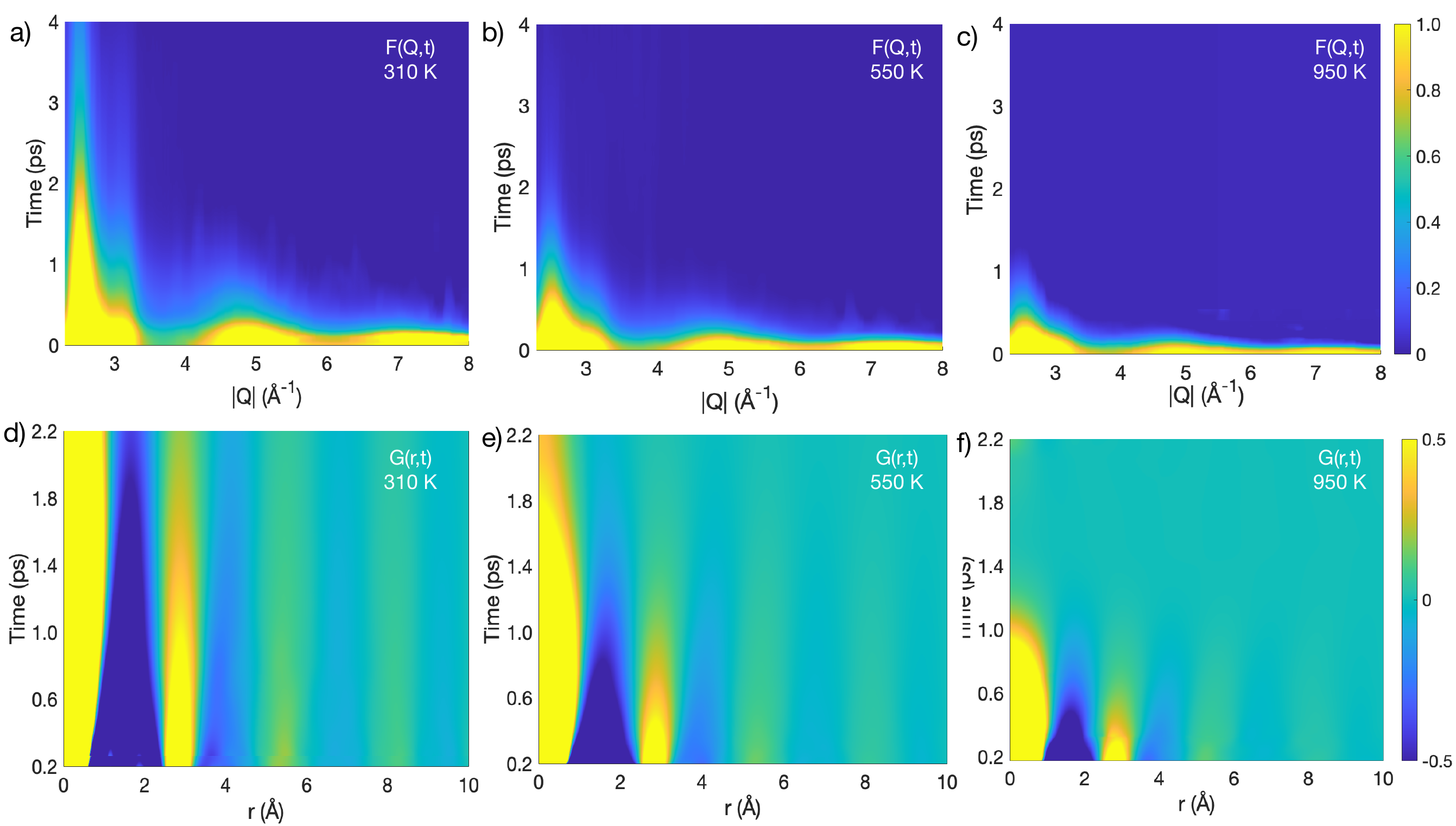}
    \caption{First row: resolution corrected Intermediate structure function, $F(Q,t)$ of liquid gallium at (a) 310 K, (b) 550 K, and (c) 950 K obtained by the Fourier transform of $S(Q,E)$ over the energy transfer, $E$; Second row: Van Hove function, $G(r,t)$, of liquid Gallium at (d) 310 K, (e) 550 K, and (f) 950 K obtained by the Fourier transform of $F(Q,t)$ over the momentum, $Q$.}
    \label{fig:Fig1}
\end{figure*}

Inelastic neutron scattering (INS) experiments on liquid gallium were conducted at the Wide-Angular Range Chopper Spectrometer (ARCS) \cite{abernathy_ARCS_2012} at the Spallation Neutron Source (SNS), Oak Ridge National Laboratory. Measurements were carried out across a wide temperature range, from 310 K to 950 K. For the measurements at 310, 350, 390, 430, 470, 510, and 550 K, 99.9\% pure gallium was loaded into a vanadium cylindrical container with an inner radius of 4 mm and a height of 60 mm. The container was sealed with a titanium lid using an indium wire as a gasket and then placed inside a MICAS furnace \cite{niedziela_MICAS_2017}.

For the measurements at 550, 650, 750, 850, and 950 K, gallium samples were first loaded into a fused silica tube with an inner radius of 3.5 mm and then placed inside a vanadium can with an inner radius of 9.5 mm and a height of 60 mm. The sample in the vanadium container was initially heated to 400 K to fully melt the gallium crystal, followed by cooling to 310 K to initiate data collection. An empty container measurement was also performed to subtract the background signal.

\textcolor{black}{For temperatures below 510 K, inelastic neutron scattering was measured using three different incident neutron energies: 20, 40, and 80,meV. At higher temperatures (above 510 K), where the signal is more strongly damped at high momentum ($Q$) and energy transfer ($E$), only two incident energies—20 and 60 meV—were used. The experiment was intentionally designed to balance energy and momentum resolution with broad $Q$–$E$ coverage by employing two to three different incident energies at each temperature. For example, the dynamic structure factor $S(Q,E)$ of liquid gallium at 310 K was measured using incident neutron energies of 20, 40, and 80 meV. At $E_i$ = 20 meV, ARCS provides an energy resolution of 0.2 meV at the elastic line and a momentum resolution of 0.1 \AA$^{-1}$, with coverage of $\pm$ 15 meV in energy transfer and 2–5 \AA$^{-1}$ in momentum transfer. At $E_i$ = 80 meV, the corresponding resolutions are 0.8 meV in energy transfer and 0.4 \AA$^{-1}$ in momentum transfer, with much broader coverage of $\pm$ 60 meV in energy transfer and 4–10 \AA$^{-1}$ in momentum transfer. Thus, low incident energy yields finer resolution, while higher energies extend the accessible Q–E range, enabling Fourier transforms to obtain $G(r, t)$. In this way, the measured $S(Q,E)$ captures the dynamics of liquid gallium on picosecond and longer timescales.}

The measured INS spectra were first reduced to the dynamic structure factor $S(Q,E)$ by using \textsc{mantid} suite \cite{arnold_mantiddata_2014}. $S(Q,E)$ spectra were then converted to the intermediate scattering function, $F(Q,t)$ by the Fourier transform over the energy transfer, $E (=\hbar\omega)$:
\begin{equation}\label{eq:SQEtoFQT}
    F(Q,t) = \int S(Q,\hbar \omega) e^{i\omega t} d\omega.
\end{equation}
The obtained $F(Q,t)$ was divided by the $F(Q,t)$ of the vanadium rod to correct for the instrumental energy resolution. \textcolor{black}{The resolution function on ARCS typically depends on both $Q$ and $E$. The $Q$-dependence usually arises from the sample geometry; however, the cylindrical geometry of our sample holder minimizes this effect. Regarding the energy dependence, calculations performed on a model dataset demonstrate that applying the resolution function at the elastic line provides an adequate correction.} Figures \ref{fig:Fig1}(a)-(c) show $F(Q,t)$ of liquid gallium at 310 K, 550 K, and 950 K after the resolution correction. Finally, the VHF, $G(r,t)$, was obtained by calculating the Fourier transform of $F(Q,t)$ over $Q$:
\begin{equation}\label{eq:FQTtoGRT}
G(r,t)-1 = \frac{\hbar}{2\rho\pi^2r}\int \text{sin}(Qr)Q F(Q,t)dQ,
\end{equation}
where $\rho$ is the atomic number density for the sample. The detailed procedure of VHF calculation is described elsewhere \cite{shinohara_real-space_2022}. The coherent neutron scattering cross-section of gallium is 7.8 barns, which is significantly higher than its incoherent scattering cross-section (0.09 barns). Therefore, the calculated VHF represents the total---self and distinct---correlation function. Results for the intermediate scattering function, $G(r,t)$, at 310 K, 550 K, and 950 K are shown in Figs.~\ref{fig:Fig1}(d)–(f).

\textcolor{black}{Kinematic restrictions limit the accessible $Q$–$E$ range, particularly at low $Q$. As a result, $S(Q)$ obtained by integrating $S(Q,E)$ over $E$ can be underestimated at low $Q$ due to the finite energy window. As noted in Ref. [3], this missing spectral weight may introduce weak, long-wavelength oscillations in the pair distribution function (PDF). In our measurements, however, excluding $Q <$ 2 \AA$^{-1}$ produces only a very weak undulation with a much longer wavelength, which does not affect the medium-range order (MRO) oscillations of interest.}

\textcolor{black}{These kinematic restrictions arise because, in inelastic scattering, $Q$ depends on both energy transfer and scattering angle, $\theta$. In diffraction experiments without energy discrimination, the intensity is integrated over energy at constant 2$\theta$ rather than at constant $Q$, which can distort results. By resolving $S(Q,E)$ as a function of both $Q$ and $E$, we integrate at constant $Q$ and thus avoid this distortion. This issue does not arise in X-ray diffraction, where the much higher incident beam energy suppresses the effect.}

\textcolor{black}{A potential source of uncertainty in $S(Q,E)$ is multiple scattering. Because multiple scattering is nearly angle-independent, its Fourier transform over $Q$ contributes primarily at the origin ($R = 0$). It therefore does not introduce oscillatory features in the PDF and does not affect the observed MRO. Another possible uncertainty is Ga–V alloying. However, its impact is negligible, as confirmed by the absence of any detectable differences in MRO when using vanadium versus silica sample containers at 550 K. This also rules out scattering from the fused silica tube as a contributor to the reported MRO features.}

The VHF, $G(r,t)$, consists of two components. At $r < 1\,\mathrm{\AA}$, the dominant contribution arises from the self-part, $G_s(r,t)$, which describes the single-atom self-motion. At larger distances, the function is governed by the distinct part, $G_d(r,t)$, which captures the collective motion between different atoms. The self-motion in liquid gallium, including diffusion, will be discussed elsewhere. In the present study, we focus on the distinct part of the VHF to elucidate the collective dynamics of gallium atoms.

\subsection{Anomalous structures and medium-range orders in liquid Ga}

\begin{figure*}[t]
    \includegraphics[scale = 0.42]{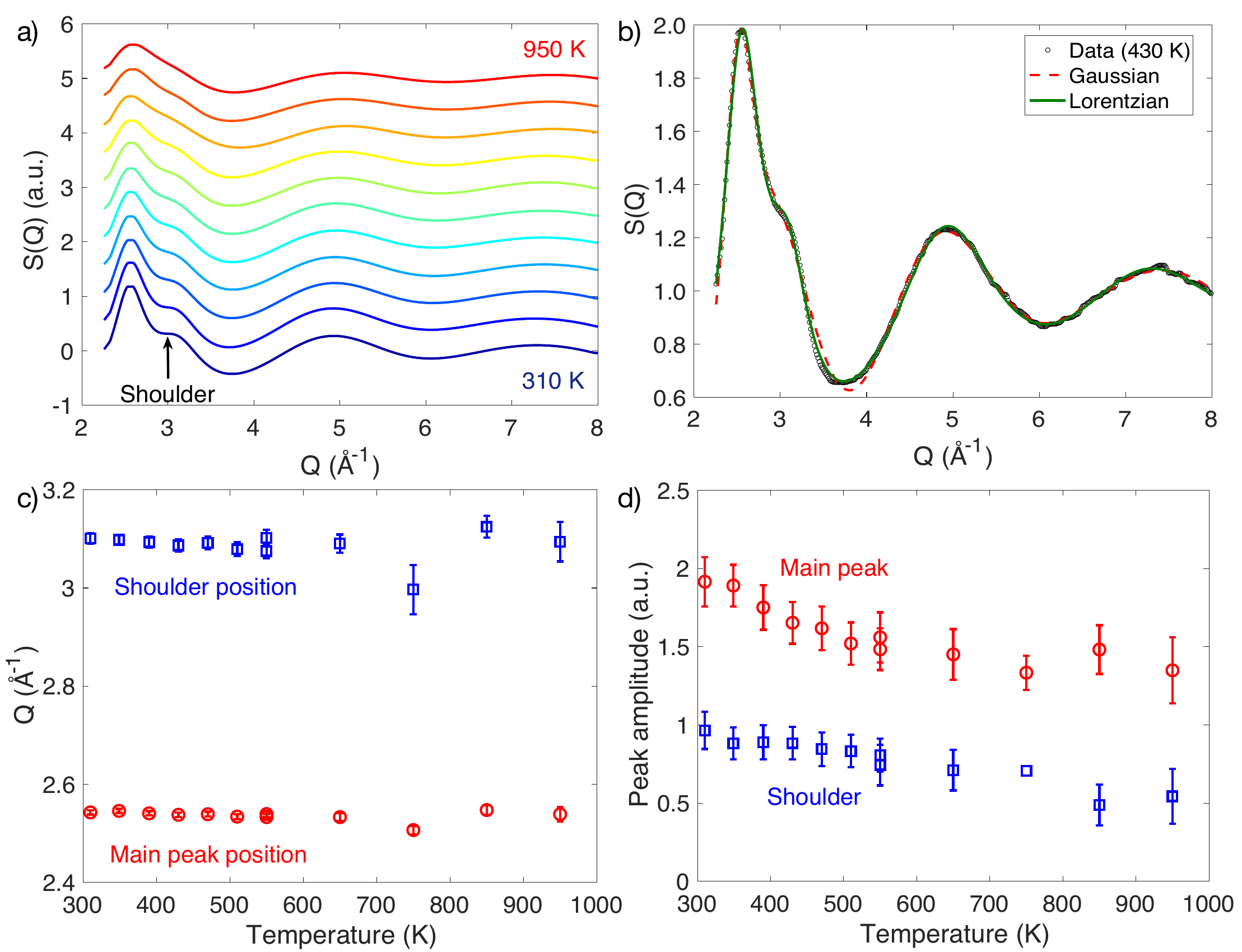}
    \caption{(a) The full set of the snapshot structure factor, $S(Q)$, from the INS measurements between 310 K (bottom) and 950 K (top). The values on the y-axis correspond to the curve at 310 K; the remaining solid curves are displaced upwards with a step of 0.5. The full set of temperatures is: 310 K, 350, 390, 430, 470, 510, 550, 650, 750, 850, and 950 K. (b) The measured $S(Q)$ at 430 K (circles) compared to the fits using a set of Lorentzian functions given by Eq.~(\ref{eq:SQ_FIT}) (solid line) and a set of Gaussian functions (dashed line). (c) The position of the main peak (circles) and shoulder (squares) of $S(Q)$ found by the Lorentzian fit given by Eq.~(\ref{eq:SQ_FIT}) as a function of temperature. (d) The corresponding amplitudes, $a_1$ and $a_2$ of the main peak (circles) and shoulder (squares) of $S(Q)$ as a function of temperature.}
    \label{fig:Fig2}
\end{figure*}

The intermediate scattering function at $t = 0$ yields $S(Q)$, the snapshot structure factor. The solid lines in Fig.~\ref{fig:Fig2}(a) show $S(Q)$ obtained from inelastic neutron scattering (INS) measurements conducted between 310 K and 950 K. A shoulder on the high-$Q$ side of the first peak is clearly observed in $S(Q)$ at 310\,K, in good agreement with structure factors obtained from X-ray and neutron diffraction \cite{narten_liquid_1972}.  As the temperature increases, this shoulder feature becomes less prominent; however, the first peak of $S(Q)$ remains asymmetric. To track the position and amplitude of the main peak and shoulder as a function of temperature, we fit the following Lorentzian functions to $S(Q)$ up to 10\,\AA$^{-1}$:
\begin{eqnarray}\label{eq:SQ_FIT}
S_{\mathrm{fit}}(Q) &=& \frac{a_1}{b_1(Q-Q_{\mathrm{main}})^2+1}+\frac{a_2}{b_2(Q-Q_{\mathrm{shoulder}})^2+1} \nonumber \\
&+&\sum^{N=4}_{i= 3}\frac{a_i}{b_i(Q-Q_i)^2+1}.
\end{eqnarray} 
The functional form of the peaks---either a Gaussian or Lorentzian function---is chosen to provide the best fit to $S(Q)$ across all temperatures. At higher temperatures, the best-fit Gaussian function does not capture the shoulder feature as effectively as the Lorentzian fit, as shown in Fig.~\ref{fig:Fig2}(b). From the Lorentzian fitting, we are able to track $Q_{\text{main}}$ and $Q_{\text{shoulder}}$ as functions of temperature, along with their corresponding peak amplitudes, $a_1$ and $a_2$, as shown in Figs.~\ref{fig:Fig2}(c) and (d). At all measured temperatures, the main peak and shoulder remain located at approximately 2.5\,\AA$^{-1}$ and 3.1\,\AA$^{-1}$, respectively. While the main peak amplitude, $a_1$, decreases with increasing temperature, the amplitude of the shoulder, $a_2$, shows only a weak temperature dependence. The weak temperature dependence of both the position and amplitude of the shoulder feature suggests that the underlying mechanism responsible for this feature persists up to high temperatures.

Many previous studies on the structure of liquid gallium have focused on the origin of this asymmetric feature at $Q_{\mathrm{shoulder}}$ in reciprocal space, interpreting this through the SRO, e.g., the coexistence of covalent and metallic characters of solid $\alpha$-Ga withing the liquid phase \cite{gong_coexistence_1993,nield_changes_1998,lambie_resolving_2024,yang_firstprinciples_2011} or the presence of locally ordered structure \cite{mokshin_short-range_2015}, but the experimental confirmation is still incomplete. Other studies have suggested that Friedel oscillations play a role in the emergence of the shoulder feature \cite{tsai_revisiting_2010, tsai_entropy_2011,hafner_structural_1990}. To investigate the physical origin of the asymmetric feature, we first aim to establish a physical connection between the snapshot structure factor, $S(Q)$, and the pair distribution function (PDF), $g(r)$. Mathematically, these two functions are related by the Fourier transformation, as expressed in Eq.~(\ref{eq:FQTtoGRT}). Often the first peak of $S(Q)$ is directly related to the first peak in $g(r)$, but such correspondence is misleading. Physically, the first peak of $S(Q)$ primarily represents the MRO, which corresponds to the higher-order peaks in $g(r)$; conversely, the first peak of $g(r)$ is mainly determined by the higher-order components of $S(Q)$ \cite{cargill_structure_1975,ryu_ideality_2020}. 

\begin{figure*}[t]
    \includegraphics[scale = 0.7]{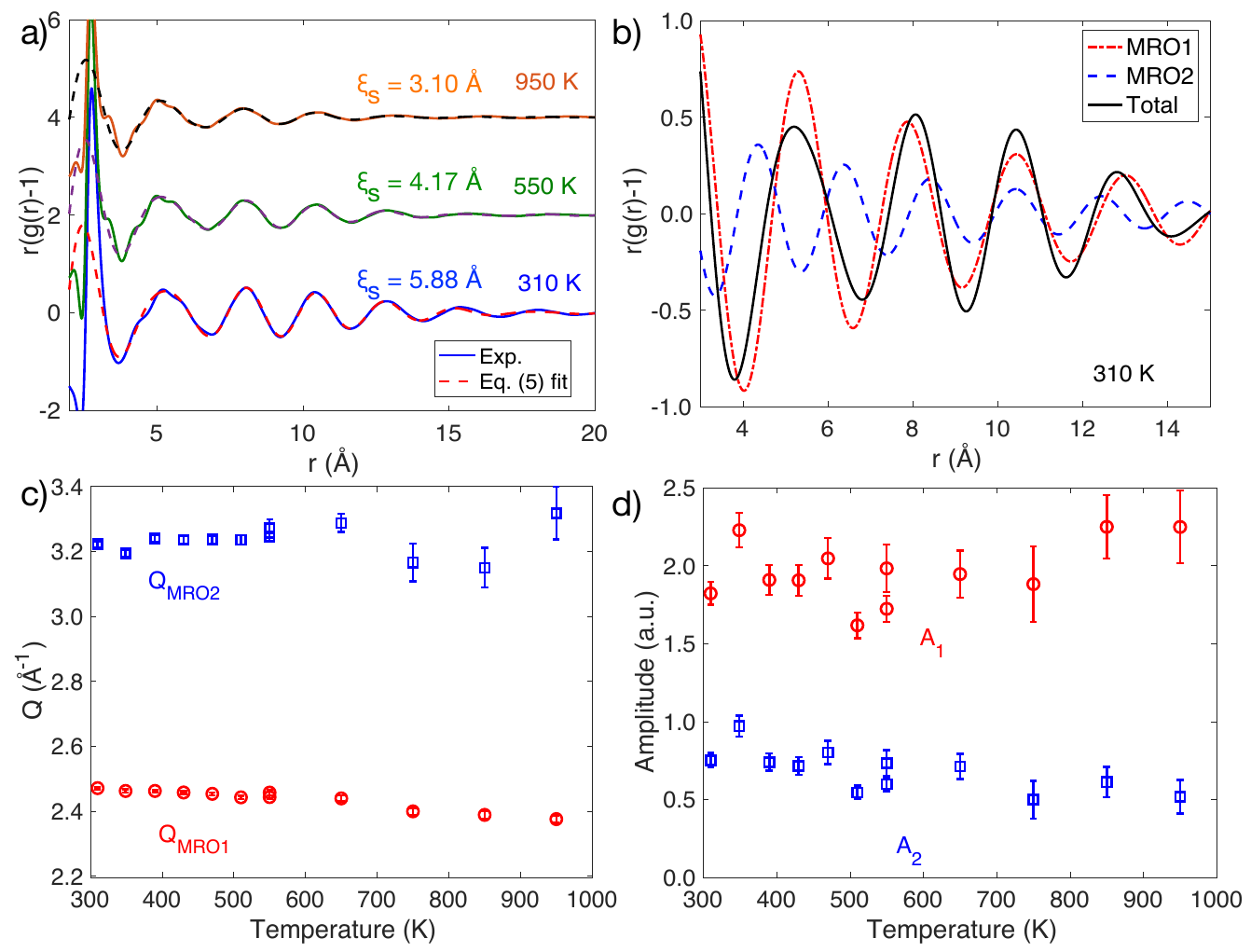}
    \caption{(a) The measured pair distribution function, $r(g(r)-1)$ (solid lines) and the best fit (dashed lines) of the higher-order peaks ($r > 3.8 \mathrm{\AA}$) using Eq.~(\ref{eq:MRO}) (dashed line) at 310 K, 550 K, and 950 K. The best-fit structural coherence length, $\xi_s$, is 5.88 $\mathrm{\AA}$ at 310 K, 4.17 $\mathrm{\AA}$ at 550 K, and 3.10 $\mathrm{\AA}$ at 950 K. (b) The best-fit $r(g_{MRO}(r)-1)$ (solid line) at 310 K consists of two decaying sinusoidal functions with wavevectors at $Q_{\mathrm{MRO1}} = 2.5\ \mathrm{\AA}^{-1}$ (dotted-dash line) and $Q_{\mathrm{MRO2}} = 3.2\ \mathrm{\AA}^{-1}$ (dashed line).  (c) Fitted medium-range order wavevectors of the main peak ($Q_{\mathrm{MRO}1}$, circles) and shoulder ($Q_{\mathrm{MRO}2}$, squares) versus temperature from the snapshot PDF. (d) Fitted amplitudes of two MRO oscillations, $A_1$, (main MRO, circles) and, $A_2$, (shoulder MRO, squares) versus temperature.}
    \label{fig:Fig3}
\end{figure*}

Since the first peak of $g(r)$ describes SRO in the nearest-neighbor shell, the presence of more than one type of local arrangement around an arbitrary atom---arising from either the mixed nature of bonding or molecular clustering---would result in a splitting or asymmetry of the first peak in $g(r)$. \textit{ Ab initio} calculations of liquid gallium predict that the coordination shells associated with covalent bonding occurs at $r \lesssim$ 2.45\,\AA{} and that with metalic bonding approximately 2.7–2.8\,\AA{}. The solid lines in Fig.~\ref{fig:Fig3}(a) show a modified form of the PDF, $r(g(r) - 1)$, at 310\,K, 550\,K, and 950\,K. Across this temperature range, $g(r)$ for liquid gallium exhibits a single, well-defined first peak, indicating that only one structural coordination shell is resolved in the measurement.

 Let us turn our attention to the higher-order peaks of $g(r)$. Recently, we proposed that, for a simple liquid, the extended oscillations in $r(g(r)-1)$ are approximately described by an exponentially decaying sinusoidal function of the form of $\sin(rQ_{\mathrm{MRO}}+\delta)\exp(-r/\xi_s)$, where $\xi_s$ is the structural coherence length defined in Ornstein-Zernike theory \cite{ryu_medium-range_2021}. $Q_{\mathrm{MRO}}$ is called the MRO wavevector, and its value is very close to the position of the first peak of S(Q). For gallium, we found that a single sinusoidal function not only provided a poor fit to the measured $g(r)$---or $G(r,t)$ in general---at all temperatures, but also failed to reproduce the shoulder feature after inverse Fourier transform back to the reciprocal space. Since two Lorentzian functions are required to model the asymmetry of the first peak in $S(Q)$, we modified the MRO equation to include two sinusoidal oscillations, given by
\begin{eqnarray}\label{eq:MRO}
 && G_{\mathrm{MRO}}(r)-1  \approx \left[A_{1}\sin(rQ_{\mathrm{MRO}1}+\delta_{1}) \right. \\ 
&& \left.+A_{2}\sin(rQ_{\mathrm{MRO}2}+\delta_{2})\right]\frac{\exp\left(-r/\xi_s\right)}{r}, \ r>r_{\mathrm{cutoff}}  \nonumber
\end{eqnarray}
where $r_{\mathrm{cutoff}}$ is the position of the first minimum of $g(r)$ beyond the first peak. The best fits using Eq.~(\ref{eq:MRO}) at 310 K, 550 K, and 950 K are shown as dashed lines in Fig.~\ref{fig:Fig3}(a). Figure~\ref{fig:Fig3}(b) shows the individual oscillations generated by the two MRO components at 310 K. The amplitudes of the two sinusoidal functions are similar, indicating that both are essential to describe the extended oscillations accurately. The relative phase difference between the two oscillations results in a weak decay of $g(r)$ between $r = $ 4 and 10 $\mathrm{\AA}$. It is found that a single coherence length, $\xi_s$, is sufficient to capture the decay behavior of both components. Figures \ref{fig:Fig3}(b) and (c) give the positions ($Q_{\mathrm{MRO1}}$ and $Q_{\mathrm{MRO2}}$) and amplitudes ($A_1$ and $A_2$) of the two MROs as a function of temperature, extracted from the fits to $G(r,t)$. The values of $Q_{\mathrm{MRO1}}$ and $Q_{\mathrm{MRO2}}$ closely match the positions of the main peak and the shoulder in $S(Q)$, respectively. Both the amplitudes and spatial frequencies exhibit weak temperature dependence, indicating that these two MROs not only coexist but also persist at high temperatures. The overall reduction in the oscillation amplitude---specifically, the amplitude ($a_1$) associated with the main peak of $S(Q)$---is attributed to a decreasing structural coherence length as temperature increases. It is important to emphasize that the asymmetric shape of the first peak in $S(Q)$ is directly linked to higher-order peaks of $g(r)$, and is therefore a manifestation of MRO rather than SRO. In the next section, we will explore this connection to MRO from a dynamical perspective.

\subsection{Collective dynamics in liquid Ga}

\begin{figure*}[t]
    \includegraphics[scale = 0.3]{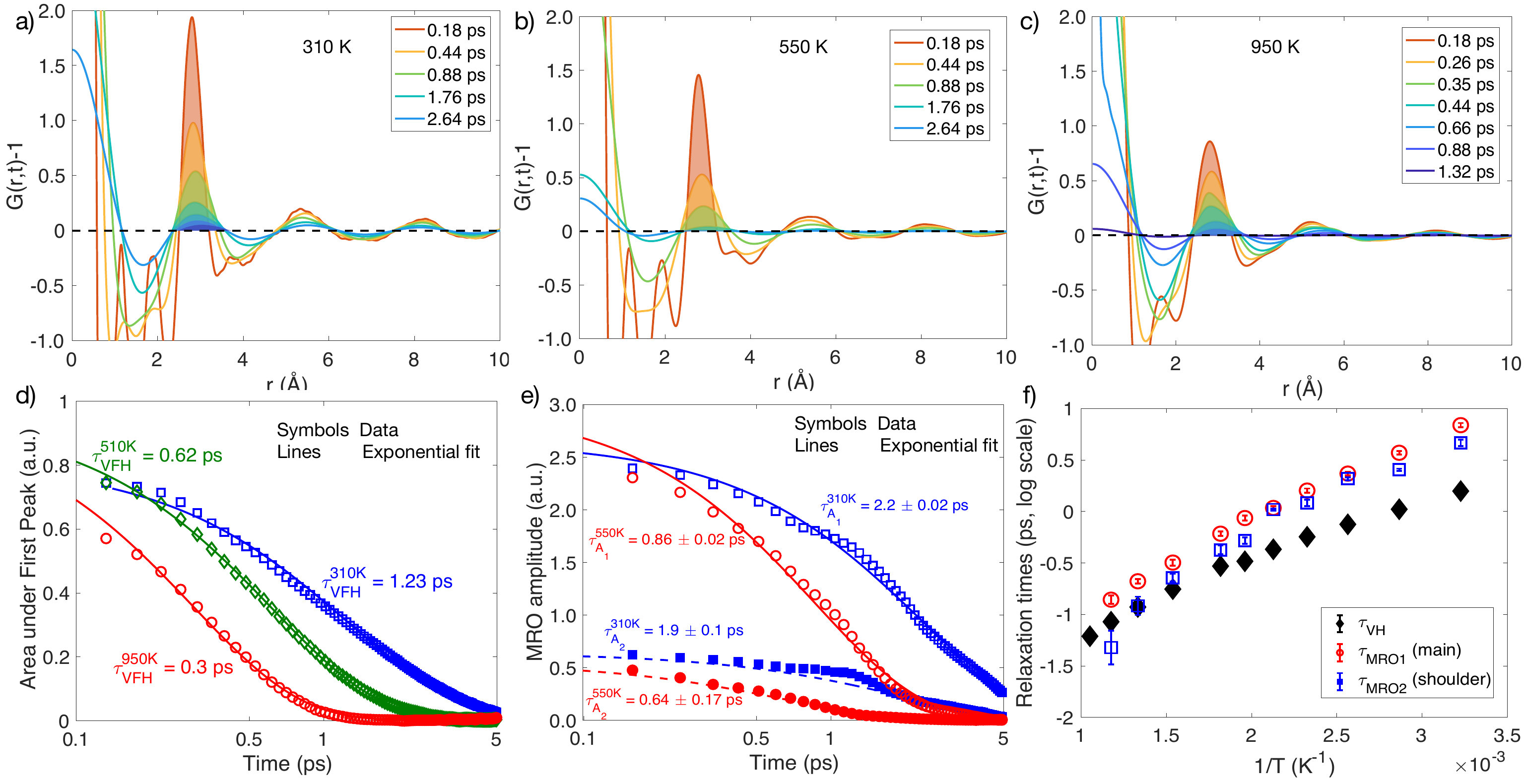}
    \caption{The time slices of VHF, $G(r,t)-1$, at (a) 310 K, (b) 550 K, and (c) 950 K. The decay behavior of the first peak describes the dynamics of the SRO while the decay behavior of the higher-order peaks gives the MRO dynamics. (d) The decay of the area under first peak of $G(r,t)-1$, $N(t)$, at 310 K (circles), 550 K (squares), and 950 K (diamonds) with their corresponding best-fit exponential decay curves. (e) The decay of  $A_1(t)$ (open symbols) and $A_2(t)$ (filled symbols) at 310 K (circles) and 550 K (squares) with their corresponding best-fit exponential decay curves. (f) Relaxation times of the first peak ($\tau_{\mathrm{VH}}$, solid diamonds), the MRO corresponding to the main peak in $F(Q,t)$ ($\tau_{\mathrm{MRO}1}$, open circles), and the shoulder MRO  ($\tau_{\mathrm{MRO}2}$, open squares), as a function of temperature. }
    \label{fig:VHF}
\end{figure*}

The distinct part of the VHF contains dynamical information about atom–atom correlations, enabling the study of the decay behavior of both first-neighbor correlations and collective atomic dynamics. Figures~\ref{fig:VHF}(a)-(c) show time slices of the VHF, $G(r,t)-1$, at 310~K, 550~K, and 950~K. The dynamics of the SRO are characterized by the temporal decay of the first peak, which corresponds to correlations at the first-neighbor correlation shell. The integrated peak intensity is computed for each temperature from
\begin{equation}\label{eq:Coordination}
N(t) = \int_D 4\pi r^2 \rho (G_d(r,t)-1)dr,
\end{equation}
where $D$ is the positive region of the first peak of the integrand shown as the shadow area in Figs.~\ref{fig:VHF}(a)-(c). \textcolor{black}{Equation (\ref{eq:Coordination}) describes how the excess density at the nearest-neighbor shell decays with time. It decays exponentially as $N_0 \text{exp}(-t/\tau_{VH})$, where $\tau_{VH}$ is referred to as the Van Hove time \cite{ashcraft_experimental_2020}.  The exponential decay has been observed in other systems (\emph{e.g.}, Refs. \cite{shinohara_real-space_2022} and \cite{shinohara_viscosity_2018}) and provides a useful means of characterizing the overall relaxation dynamics.} Figure \ref{fig:VHF}(d) shows $N(t)$ at three temperatures, along with their corresponding best-fit exponential decay curves. The Van Hove time characterizes the timescale over which an atom loses certain number of nearest neighbors, to be replaced by new ones. This timescale typically depends on both the distance of an atom from the central atom’s initial position and the local structural environment \cite{ashcraft_experimental_2020}.

The dynamics of the two MROs are studied by fitting the higher-order peaks of $G_d(r,t)$ using time-dependent parameters $A_1(t)$, $A_2(t)$, $\delta_1(t)$, and $\delta_2(t)$, while keeping $Q_{\mathrm{MRO1}}$, $Q_{\mathrm{MRO2}}$, and $\xi_s$, fixed at their values at $t=0$. The fitted values of $A_1(t)$ and $A_2(t)$ as functions of time at 310 K and 550 K are shown in Fig.~\ref{fig:VHF}(e). The decay of $A_1(t)$ and $A_2(t)$ can be described by exponential functions, $A_1(0)\exp(-t/\tau_{\mathrm{MRO1}})$ and $A_2(0)\exp(-t/\tau_{\mathrm{MRO2}})$, where $\tau_{\mathrm{MRO1}}$ and $\tau_{\mathrm{MRO2}}$ are defined as the relaxation times of the two MROs \cite{shinohara_viscosity_2018}. Figure~\ref{fig:VHF}(f) shows $\tau_{\mathrm{VH}}$, $\tau_{\mathrm{MRO1}}$ and $\tau_{\mathrm{MRO2}}$ as a function of inverse temperature.  The relaxation times for the two MROs remain similar across all temperatures, and their decay is attributed to thermal density fluctuations \cite{Ryu_compositional_2025}. Another notable feature in Fig.~\ref{fig:VHF}(f) is that, at low temperatures, $\tau_{\mathrm{VH}}$ is significantly smaller than both $\tau_{\mathrm{MRO1}}$ and $\tau_{\mathrm{MRO2}}$, indicating that SRO relaxes much faster than MRO \cite{wu_atomic_2018}. This experimental observation challenges the conventional view that MRO emerges solely as a consequence of SRO. Notably, MRO persists well beyond the complete decay of SRO, suggesting that it may originate from an independent mechanism or underlying structural process.

\textcolor{black}{It is important to note that the relaxation process of each MRO occurs at a time scale (t $>$ 1 ps), significantly longer than the fast regime (t $<$ 1 ps) typically governed by phonon-like excitations \cite{PhysRevLett.89.255506,PhysRevB.71.014207,khusnutdinoff_collective_2020}. In contrast, the timescale for structural relaxation we focus here is longer (t $>$ 1 ps)}

\begin{figure*}[t]
    \includegraphics[scale = 0.45]{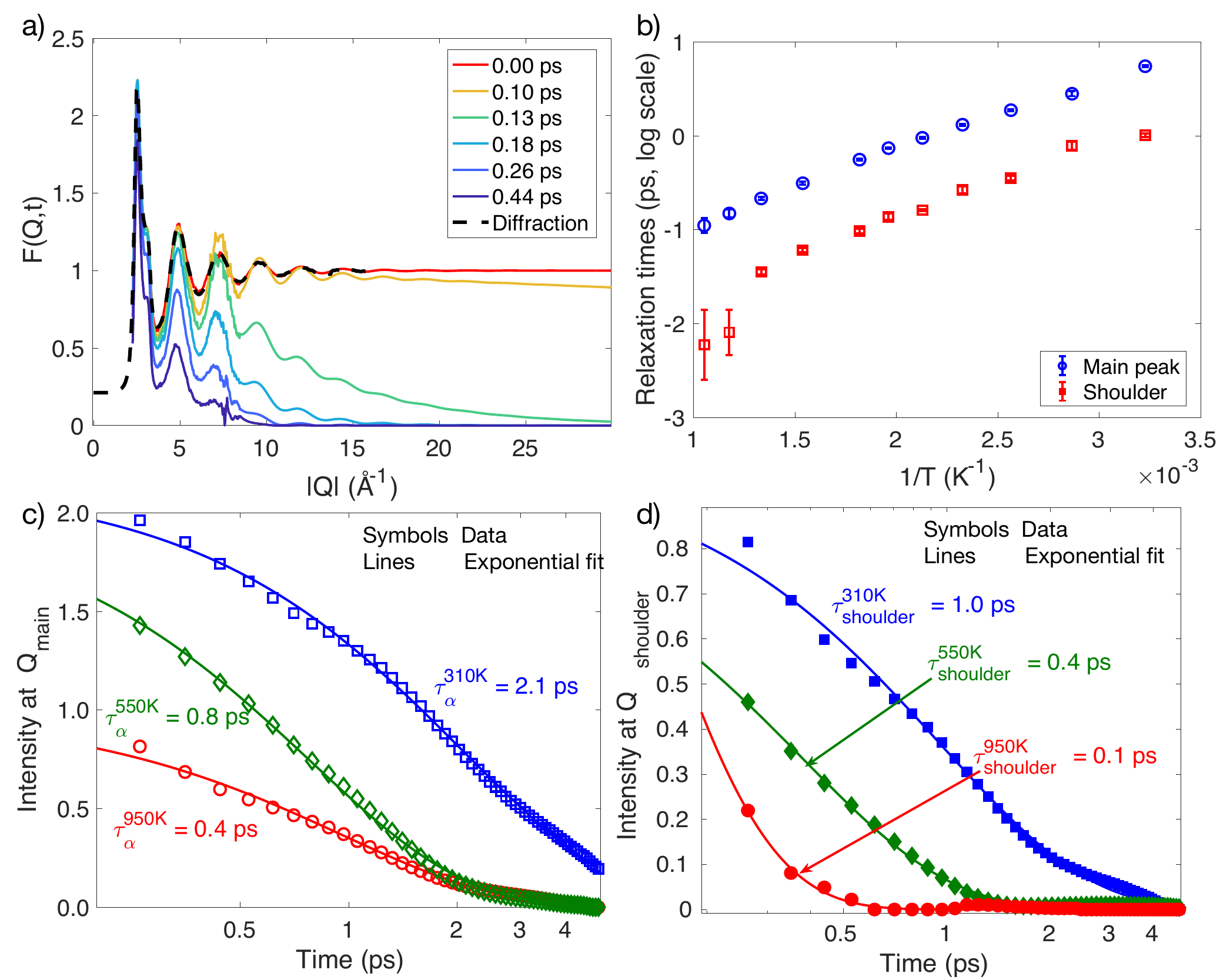}
    \caption{(a) The time slices of the intermediate scattering function, $F(Q,t)$, at 310 K. The dashed line is the x-ray diffraction measurement from Ref. \cite{narten_liquid_1972}. (b) Best-fit relaxation times of the amplitude decay at $Q = Q_{\mathrm{main}}$ and $Q = Q_{\mathrm{shoulder}}$ as a function of inverse temperature. (c) The amplitude decay of F(Q,t) at $Q = Q_{\mathrm{main}}$ at 310 K (squares), 550 K (diamonds), and 950 K (circles) with their corresponding best-fit exponential decay curves. (d) The amplitude decay of F(Q,t) at $Q = Q_{\mathrm{shoulder}}$ at 310 K (squares), 550 K (diamonds), and 950 K (circles)  with their corresponding best-fit exponential decay curves.}
    \label{fig:FQT}
\end{figure*}

Before concluding this section and beginning the discussion on the origin of the two types of MRO, it is important to emphasize that the dynamics of MROs are unambiguously accessible through the measurements of real-space correlation functions. The results of inelastic scattering measurements are usually expressed in terms of the intermediate scattering function, $F(Q,t)$ to extract dynamic parameters such as diffusion coefficients or relaxation times. However, its decay beyond the first peak typically reflects a mixture of self-motion and correlated motion. For example, Fig.~\ref{fig:FQT}(a) shows the time slices of $F(Q,t)$ at 310 K.  The amplidue decays rapidly with $Q$ particularly at longer times, indicating contributions from both self and collective motion.  Figures \ref{fig:FQT}(c) \& (d) give the amplitude decay at $Q = Q_{\mathrm{main}}$ and $Q = Q_{\mathrm{shoulder}}$ at 310 K, 550 K, and 950 K.  Both decays can be described well by exponential functions, $S(Q_{\mathrm{main}}) \text{exp}(-t/\tau_{\alpha})$ and $S(Q_{\mathrm{shoulder}}) \text{exp}(-t/\tau_{\mathrm{shoulder}})$. The best-fit relaxation times,  $\tau_{\alpha}$ and $\tau_{\mathrm{shoulder}}$, as a function of inverse temperature are plotted in Fig.\ref{fig:FQT}(b). Notably, the MRO relaxation times and $\tau_\alpha$ are nearly identical across all temperature, reinforcing that the first peak of $F(Q,t)$ predominantly reflects the high-order peaks of $G(r,t)$, and thus the MRO. 

In contrast,  $\tau_{\mathrm{shoulder}}$ in Fig.~\ref{fig:FQT}(b) is much smaller than that of the main peak. A direct comparison between the two values would be misleading, because the shoulder of the first peak in $F(Q,t)$ is heavily influenced by the decay of the self-correlated term associated with diffusion. Therefore, $\tau_{\mathrm{shoulder}}$ should not be directly compared to the relaxation times of MRO2. This underscores the limitation of interpreting liquid dynamics solely through $F(Q,t)$ in reciprocal space: there is no straightforward one-to-one correspondence between the features in real-space and those in $Q$-space. On the other hand, by Fourier transforming $F(Q,t)$ to $G(r,t)$, we can disentangle the decay behavior of self-correlation at short distances from those of interatomic correlation at longer distances, providing a clearer picture of the underlying dynamics.

\section{Discussion}

To identify the origin of these two MROs, it is important to recognize that MRO is fundamentally different in nature from SRO. As shown in the above analysis, the dynamics of medium-range correlations---those extending beyond the first nearest neighbors---are clearly distinct from those of the short-range structure represented by the first coordination shell. Whereas short-range order reflects local chemical bonding and atomic packing, the medium-range component of the VHF, as described by Eq.~(\ref{eq:MRO}), captures coarse-grained density fluctuations, as each peak covers hundreds and thousands of interatomic distances \cite{egami_local_2020}. The extended oscillations in $G_d(r,t)$ reflect modulations in the probability distribution of interatomic separations, suggestive of an emergent density wave (DW)–like ordering. Extrapolating the correlation length $\xi_s$ in Eq.~\ref{eq:MRO} to infinity yields a hypothetical state characterized by two distinct long-range DWs \cite{ryu_curie-weiss_2019}. However, such a state is intrinsically unstable against local thermal structural fluctuations, which limit the spatial extension of the DWs and result in the emergence of exponentially damped MRO \cite{egami_medium-range_2021}. In this framework, MRO reflects the finite strength and coherence of these damped DW states and represents a cooperative, mesoscopic level of structural organization—beyond the atomic-scale features governed by SRO \cite{egami_medium-range_2023}.

\textcolor{black}{The DW theory has been widely used in condensed-matter physics for a long time \cite{PhysRevLett.41.702, PhysRevB.32.4592}. It is an alternative way to describe the complex structure, and the DWs are just the Fourier-components of the atomic density, $\rho(\textbf{r})$. Therefore, usually they are static waves. In liquids, they describe dynamic structure and do not include phonons, which are vibrations at shorter timescales. The temporal and spatial decay of density waves occurs through local atomic rearrangements due to thermal fluctuations of local structures \cite{egami_medium-range_2021}. Consequently, even though the two MROs arise from different microscopic origins, they display the identical spatial and temporal decay behaviors, governed by the same structural coherence length and relaxation time.}

\begin{figure*}[t]
    \includegraphics[scale = 0.42]{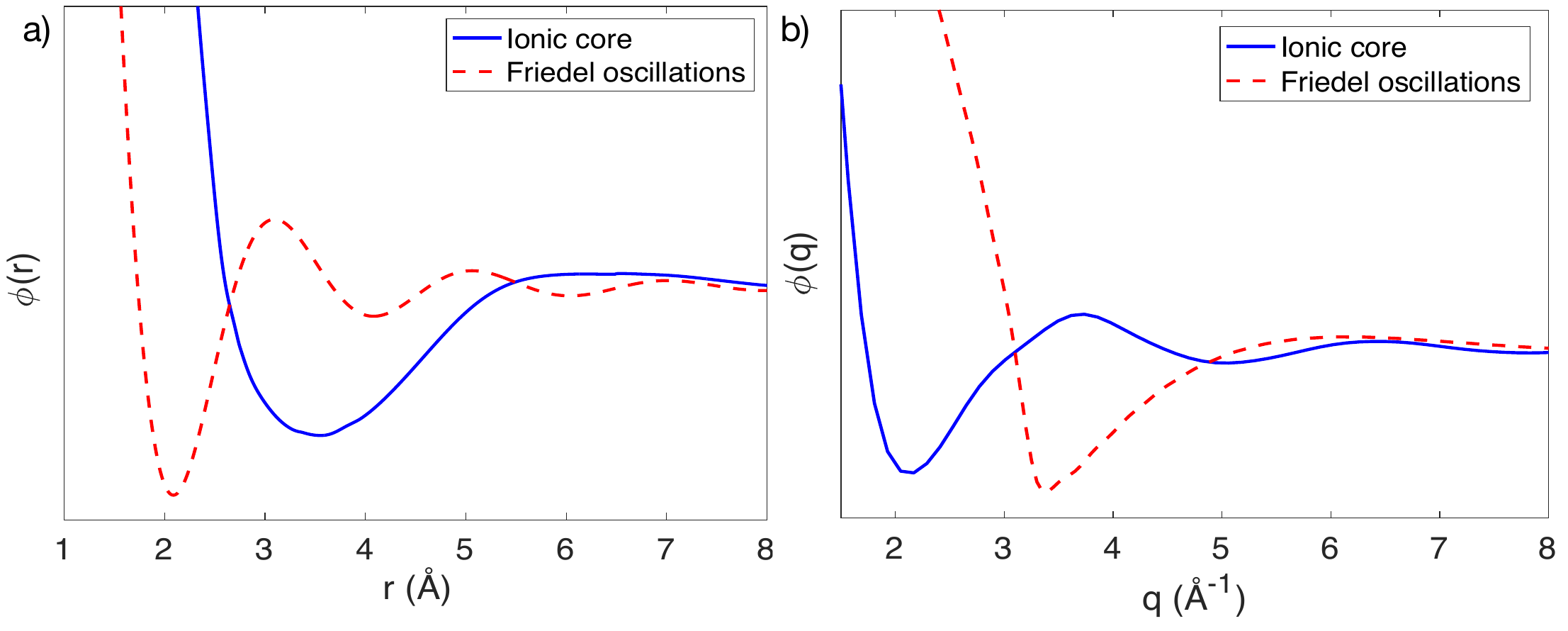}
    \caption{(a) Pseudopotential, $\phi_{pp}(r)$, in liquid Gallium consists of an interatomic pair potential for the ionic component (solid line, Ref.~\cite{belashchenko_computer_2012}) and long-range Friedel oscillations induced by conduction electrons (dashed line). (b) Pseudopotential, $\phi_{pp}(q)$ expressed in $q$-space, shows deep minima at $q_1 \approx 2.2~\mathrm{\AA}^{-1}$ for the ionic pair potential and at $q_2 \approx 3.2~\mathrm{\AA}^{-1}$ for Friedel oscillations.}
    \label{fig:Dynamics}
\end{figure*}

What are the driving forces behind the two DWs observed in liquid gallium? One promising approach to addressing this question is to consider the interatomic potential in reciprocal space \cite{egami_structural_2022}. This formalism enables a collective treatment of the interactions among all atoms, capturing the cooperative effects of the potential. For a simple liquid with a spherically symmetric interatomic potential, $\phi(r)$, the total potential energy can be expressed in reciprocal space as:
\begin{equation}
U = \int \rho(\textbf{r})\rho^*(\textbf{r}')\phi(|\textbf{r}-\textbf{r}'|)d\textbf{r}d\textbf{r}' = \int |\rho(\textbf{q})|^2\phi(|\textbf{q}|)d\textbf{q},
\end{equation}
where $\rho(\textbf{r})$ is the local density and related to the structure function, $S(\textbf{q})$, by $S(\textbf{q}) \propto |\rho(\textbf{q})|^2$.  $\phi(r)$ can be divided into two parts: a highly repulsive component $\phi_R(r)$, which dominates for $r<r_c$, and a pseudopotential component $\phi_{pp}$, which applies for $r > r_c$. The cutoff distance $r_c$ is chosen such that the corresponding cutoff energy, $k_BT_u = \phi_R(r_c)$, is well above the actual thermal energy. This ensures that virtually no pair of atoms is found at distances shorter than $r_c$. Then the total potential energy of the system can be rewritten as $U = U_{pp} + U_R$. The ground-state structure is determined by minimizing the total energy $U$. Because no pair of atoms is found at $r < r_c$, $U_R$ is zero, and the structural configuration is entirely determined by minimizing $U_{pp}$. 

Within this framework, the central question becomes: what constitutes the pseudopotential $\phi_{pp}$ in liquid gallium? Oberle and Beck \cite{oberle_influence_1979} and others \cite{tsai_revisiting_2010,tsai_entropy_2011,hafner_structural_1990} have suggested that the pseudopotential in polyvalent metal melts comprises two main contributions: an interatomic pair potential associated with the ionic core and Friedel oscillations induced by conduction electrons \cite{schommers_pair_1983,faber_introduction_2010}. For liquid gallium, the repulsive core of the pair potential exhibits a minimum at approximately 3.6 $\mathrm{\AA}$ \cite{belashchenko_computer_2012}. The Friedel oscillatory component of the potential can be described by $\phi_{\mathrm{FO}} \propto \cos(2k_{\mathrm{F}} r+\alpha)/(2k_{\mathrm{F}}r)^3$, where $k_{\mathrm{F}} = 1.624\ \mathrm{\AA}$ is the Fermi wavevector in liquid gallium, obtained from first principle calculations \cite{tsai_revisiting_2010}. Figure~\ref{fig:Dynamics}(a) illustrates the two contributions to the real-space pseudopotential of liquid gallium. Their relative amplitudes are undetermined, since a more detailed analysis is required to determine them. Here, the focus here is to highlight the qualitative shape and spatial extent of the potential components. These two potentials influence atomic behavior in fundamentally different ways. The core potential acts directly as a two-body interaction, exerting forces between neighboring atoms. In contrast, the Friedel potential operates indirectly: an ion induces local charge density waves (CDW) in the valence electron distribution, which in turn mediates interactions with nearby ions. This distinction is analogous to the difference between the direct exchange interactions and indirect Ruderman–Kittel–Kasuya–Yosida (RKKY) interactions in magnetic metals. Accordingly, it is reasonable to treat the effects of these two potentials separately. When expressed in reciprocal space, these two potentials exhibit distinct minima: the ionic component has a minimum at $q_1 \approx 2.2\ \mathrm{\AA}^{-1}$ for the core repulsion, while the Friedel oscillatory component shows a minimum at $q_2 = 2k_F \approx 3.2\ \mathrm{\AA}^{-1}$ (Fig.~\ref{fig:Dynamics}(b)). In our view, it is not a coincidence that $q_1$ and $q_2$ are in close proximity to $Q_{\mathrm{MRO}1}$ and  $Q_{\mathrm{MRO}2}$, respectively. We propose that these two minima in the pseudopotentials provide the driving forces for the two DWs in liquid gallium. 

If $\phi_{pp}(q)$ has a minimum at $q_{\text{min}}$, then $U_{pp}$---and consequently the total potential energy $U$---is approximately minimized when the density distribution in reciprocal space forms a density wave with $\lvert \textbf{q}_{\text{DW}} \rvert = q_{\text{min}}$, given by $\rho_{\text{DW}}(q) = \rho_{\text{DW}} \delta(q-q_{\text{DW}})$, where $\rho_{\text{DW}}$ is the amplitude of the density wave. Because $\textbf{q}_{\text{DW}}$ is densely distributed over the sphere in $q$ space with radius of $q_{\text{min}}$, this DW state includes a very large number of DWs \cite{ryu_curie-weiss_2019}. \textcolor{black}{This corresponds to a long-range, spherically symmetric density wave state characterized by the wavevector $q_{\text{min}}$, which emerges at the ideal ground state—a state with long-range correlation without lattice periodicity \cite{ryu_curie-weiss_2019}.} However, due to thermal fluctuations and geometrical frustration, structural coherence cannot fully develop in real liquid and glass. As a result, the density wave oscillations in the PDF are exponentially damped, indicating limited spatial coherence, as described by Eq.~(\ref{eq:MRO}). In liquid gallium, there are two driving forces represented by two pseudopotentials with two distinct minima. In crystals, these two driving forces must be reconciled, resulting in a single, often complex, lattice structure constrained by the requirement of translational symmetry. In contrast, liquids are not subject to such symmetry constraints, allowing the two driving forces to independently give rise to distinct DWs. We propose that the superposition of these DWs leads to the final structural configuration of liquid gallium. Note that for each kind of $\text{q}_{\text{min}}$ , there are many DWs propagating in different directions. As a result, each Ga atom experiences a superposition of numerous fluctuating DWs.

\textcolor{black}{In contrast, the first neighbor shell in the PDF is determined primarily by the potential in real space, whereas the MRO is determined by the pseudopotential in q space. Then, as shown in Fig. 6 (a), the minimum in the electronic potential in real space is overwhelmed by the strong repulsive part of the ionic potential. Consequently, the local environment around each atom changes rapidly, on a timescale much shorter than the decay time of the DWs.  This explains the short bond lifetime \cite{lambie_resolving_2024} and why only a single type of atomic environment is observed, resulting in an unsplit first peak in $g(r)$.}

The success of this framework provides further support for the density wave theory of liquids \cite{egami_origin_2023}. This interpretation of the MRO fundamentally departs from the picture proposed by Hafner and colleagues, in which MRO arising from Friedel oscillations was viewed as a spatial extension of SRO \cite{jank_structural_1990}.  In contrast, the current approach begins with a high-density gas state and derives structural organization by minimizing the global potential energy through the formation of spatially constrained density waves. Within this DW framework, the first MRO at $Q_{\mathrm{MRO1}}$ is interpreted as a direct consequence of coarse-grained density fluctuation modulation due to ion-ion repulsion. The second MRO at $Q_{\mathrm{MRO2}}$, on the other hand, is driven by electronic charge density modulations caused by electron gas near the Fermi level scattered by the ions. 

\textcolor{black}{It should be emphasized that the second MRO does not directly arise from Friedel oscillations. Although the characteristic spatial frequency of the MRO coincides with 2$k_{\mathrm{F}}$, this emerged as an outcome of our real-space analysis of G(r), not as an assumption. As two distinct phenomena, Friedel oscillations and the second MRO originate from the same underlying cause—the electronic structure of gallium metal. Importantly, while MRO oscillations may appear superficially similar to Friedel oscillations, there is a crucial distinction between them: Friedel oscillations decay as $1/r^3$ with distance, whereas MRO oscillations decay as $1/r$, making them far more extended. This difference arises from the distinct nature of the underlying phenomena—Friedel oscillations represent a local screening response to a point charge, while MRO oscillations correspond to a collective density-wave response involving many charge centers randomly distributed in space. Thus, MRO is not simply a direct consequence of Friedel contributions to the interatomic potential, as is often assumed, but rather a more collective electronic response better understood in terms of the virtual formation of charge-density waves.}

\textcolor{black}{Metals share similar electronic structure that gives rise to Friedel oscillations}, and similar structural anomalies---a split first peak in $S(Q)$---have been observed in other polyvalent metal melts such as Si, Ge, Sn, As, Sb, and Bi, indicating that the coexistence of two distinct MRO features may be a rather universal characteristics of this class of materials. In contrast, other metals such as Mg, Au, and Zn do not exhibit a shoulder or asymmetry near $Q_{\mathrm{shoulder}}$ in $S(Q)$. We suggest that the underlying competition between the repulsive ionic core potential and the cohesive interactions mediated by valence electrons determines whether complex structural features are manifested in $S(Q)$. 

Two primary factors should be considered in this competition. The first is the relative positioning of $Q_{\mathrm{MRO}1}$ and $Q_{\mathrm{MRO}2}$, which are governed by the average interatomic distance and the Fermi wavevector in liquid metals. For metals such as Mg, Au, and Zn---with a valence-electron-per-atom ratio of 2, the principle peak position, $Q_{\mathrm{main}}$ in $S(Q)$, typically aligns with $2 k_F$. This implies that the wavelength of the long-range Friedel oscillations in the interatomic potential matches the average interatomic spacing in these systems \cite{hafner_atomic_1988}, resulting in a single dominant structural feature. In contrast, for metals and metalloids in groups IIIA, IVA, VA, and VIA, which have valence-electron-per-atom ratios greater than 2, $2 k_F > Q_p$. In these cases, $Q_{\mathrm{MRO}1}$ and $Q_{\mathrm{MRO}2}$ become sufficiently distinct to appear as a split first peak in $S(Q)$. 

The second factor is the relative amplitude of the Friedel oscillations compared to the strength of the core repulsion. The relative strength between the two components of the interatomic potential governs the characteristics of the first peak in $S(Q)$: total absence of the shoulder feature , two distinct peaks, or an asymmetric peak.  For example, the near-absence of such anomalies in liquid aluminum may be attributed to a dominant core repulsion that overwhelms weak electron-mediated modulations due to the electronic pseudopotential which is accidentally shallow  \cite{oberle_influence_1979, hafner_hamiltonians_1987}. While further investigation is needed to generalize this idea beyond gallium, this framework provides a foundation for developing a more realistic and comprehensive theory of structural anomalies in liquid metals.

\section{Conclusion}
Atomic-scale dynamics in liquid gallium were investigated using inelastic neutron scattering. The total VHF was obtained by Fourier transforming the inelastic neutron scattering spectra. By analyzing the snapshot PDFs from 310 K to 950 K, we identified the coexistence of two MRO components in liquid gallium, extending beyond the first nearest neighbors. These two MROs persist up to high temperatures. The changes in the asymmetric shape of the first peak in $S(Q)$ with temperature is purely due to the change in its width, caused by the change in the MRO coherence length. 

Analysis of the measured VHFs reveals that the dynamics of two MROs are essentially identical, even though the intermediate function, $F(Q,t)$, suggests the decay times of the main peak and shoulder are different by an order of magnitude. This is because in $F(Q,t)$ the self- and distinct- correlations are mixed, whereas in VHF they are clearly separated. The decay of $F(Q,t)$ at the shoulder of the first peak is heavily influenced by the self-term, and does not faithfully describe the decay of the local structure. 

To explain the origin of two types of MRO, we propose a top-down framework based on DW theory. In this approach, MRO arises from many-body interactions governed by the pseudopotentials, which promotes the formation of density modulations in the liquid. In liquid metals such as gallium, the repulsive force from the ionic core and the extended Friedel oscillations of the conduction electrons act as two competing driving forces, leading to the formation of two types of MROs with distinct wavevectors. Despite their different origins, these MROs share the same coherence length and decay time. Each atom experiences both types of DWs, causing its local bonding character—more metallic or more covalent—to fluctuate rapidly over time. As a result, the first peak of the PDF appears as a single component. However, at the length scale associated with MRO, two distinct types of correlations emerge, driven by the two different interactions. This interpretation presents a significant challenge to the conventional understanding of this anomalous liquid

Further investigation is needed to determine whether this mechanism applies to other metallic melts. Nevertheless, the insight provided by the density wave approach offers a promising direction for future research, linking electronic structure to atomic dynamics and shedding light on the dynamic nature of liquids and glasses.

\begin{acknowledgments}
This work was supported by U.S. Department of Energy (DOE), Office of Science, Office of Basic Energy Science (BES), Division of Materials Sciences and Engineering. A portion of this research used the resources at the Spallation Neutron Source, supported by DOE, BES, Scientific User Facilities Division.The beamtime was allocated to ARCS on proposal number IPTS-21883. TE is grateful to Martin Stiehler for useful discussion.
\end{acknowledgments}

\section*{Data Availability}

The data that support the findings of this article are openly available \cite{hua_data_2025}.

%using BibTeX:
%\bibliography{MyRef}


\end{document}